\documentstyle[aps,prl,multicol,epsfig]{revtex}
\begin{document}
\draft
\title{A New Measurement of the $^1S_0$ Neutron-Neutron Scattering 
Length using the Neutron-Proton Scattering Length as a Standard}

\author{ D.E. Gonz\'{a}lez Trotter$^1$, F.~Salinas$^1$, Q.~Chen$^1$, 
A.S.~Crowell$^1$, W.~Gl\"ockle$^2$, C.R.~Howell$^1$, C.D.~Roper$^1$, 
D.~Schmidt$^3$, I.~\v{S}laus$^4$, H.~Tang$^5$, W.~Tornow$^1$, R.L. Walter$^1$, 
H.~Wita{\l}a$^6$ and Z. Zhou$^5$}

\address{$^1$Duke University and Triangle Universities Nuclear Laboratory, 
 Durham, NC 27708, USA}
\address{$^2$Institut f\"ur Theoretische Physik, Ruhr-Universit\"at Bochum, 
  D-44780 Bochum, Germany}
\address{$^3$Physikalisch-Technische Bundesanstalt, D-38116 Braunschweig, 
 Germany}
\address{$^4$Rudjer Boskovic Institute, Zagreb, Croatia.}
\address{$^5$Chinese Institute for Atomic Energy, Beijing, Peoples Republic 
 of China}
\address{$^6$Institute of Physics, Jagellonian University, PL-30059 
 Cracow, Poland}

\date{\today}

\maketitle

\begin{abstract}
The present paper reports high-accuracy cross-section data for
the $^2{\rm H}(n,nnp)$ reaction in the neutron-proton ({\em np}) and 
neutron-neutron ({\em nn}) final-state-interaction (FSI) regions
at an incident mean neutron energy of 13.0~MeV.
These data were analyzed with rigorous three-nucleon calculations 
 to determine the $^1S_0$ {\em np} and {\em nn} scattering lengths, $a_{np}$
and $a_{nn}$.  Our results are $a_{nn}$ = -18.7 $\pm$0.6~fm and 
$a_{np}$ = -23.5 $\pm$0.8~fm. 
Since our value for $a_{np}$ obtained from neutron-deuteron ({\em nd})
breakup 
agrees with that from free {\em np} scattering, we conclude that our
investigation of the {\em nn} FSI done simultaneously and under identical 
conditions
gives the correct value for $a_{nn}$.  Our value for $a_{nn}$ is in 
agreement with that obtained in $\pi^-d$ measurements but disagrees with 
values obtained
from earlier {\em nd} breakup studies.
\end{abstract}


\begin{multicols}{2}

The difference in the 
$^1S_0$ neutron-neutron ({\em nn}) and proton-proton ({\em pp})
scattering lengths is 
an explicit measure of charge-symmetry breaking ({\em CSB}) of the
the nuclear force.
The high sensitivity of these scattering lengths to 
the nuclear potential
strength makes them a valuable probe for 
detecting small potential-energy contributions like the
isospin-dependent forces  \cite{Sla89,Mil90,Wir95} that cause CSB.  
For realistic 
potentials, a 1\% change in the potential strength results in a 30\% 
shift in the scattering length. 
Since it is not technically viable to measure
$a_{nn}$ directly by conducting free {\em nn} scattering experiments, 
one uses few-nucleon reactions that emit two neutrons with low 
relative momentum,
i.e., a final-state interaction (FSI) configuration.  
The two reactions used to measure $a_{nn}$ that have the smallest 
theoretical uncertainties 
are pion-deuteron capture ($\pi^- + d \rightarrow n + n + \gamma$)
and neutron-deuteron breakup ($n + d \rightarrow n + n + p$).  
Curiously, measurements using these two reactions give significantly 
different values of $a_{nn}$:  from $\pi^-d$ capture measurements the average
value for $a_{nn}$ is $-18.6 \pm 0.4$ fm \cite{Gab79,Sch87,How98} and from 
kinematically-complete {\em nd} breakup experiments the average value is
$-16.7 \pm 0.5$ fm \cite{Sla89}.  
It was suggested that the difference in these values 
has its origin in the 
three-nucleon force (3NF), which would act 
in the {\em nd} breakup reaction \cite{Sla89} but not in the other. 

The goal of the present work was to measure $a_{nn}$ with the {\em nd}
breakup reaction to an accuracy better than $\pm$0.7~fm.  
To this aim $a_{np}$ was measured simultaneously and used
as a standard for evaluating our methods and for determining 
the influence of the 3NF in {\em nd} breakup. 
We obtained $a_{np}$ and $a_{nn}$ from absolute cross-section measurements 
in the {\em np} and {\em nn} FSI regions, respectively, in {\em nd} breakup 
at an incident mean neutron energy of $E_{n}=13.0$~MeV. 
Rigorous {\em nd} breakup calculations with the 
Tucson-Melbourne (TM) 3NF potential~\cite{tm3nf} show
that the percentage change in {\em np} and {\em nn} FSI cross 
sections from the addition of a 3NF are equal.  
Based on this 
observation and the fact that $a_{np}$ has been determined to high
accuracy by {\em np} scattering~\cite{Koe75}, we used our extracted values of
$a_{np}$ to set an upper limit
on the 3NF influence on the value of $a_{nn}$ determined 
in the present experiment.
Because 3NF effects might have a stronger energy and angle dependence than 
indicated by our calculations with the TM 3NF model, 
we designed the experiment to obtain {\em nn} FSI data at the same
energy and angle configurations as that for the {\em np} FSI
data.  Furthermore, guided by our calculations of the angle dependence 
of 3NF effects 
on the {\em NN} FSI cross-section enhancement\cite{wit95},  
measurements were made at several {\em np} and {\em nn} emission angles.
The measured cross-section distribution at each {\em np} ({\em nn}) angle 
was used 
to determine a value of $a_{np}$ ($a_{nn}$), and the predicted angle dependence
of $a_{np}$ ($a_{nn}$) due to 3NF effects 
was investigated.  

\begin{figure}
 \noindent
 \begin{minipage}{\linewidth}
  \begin{center}
    \epsfig{file=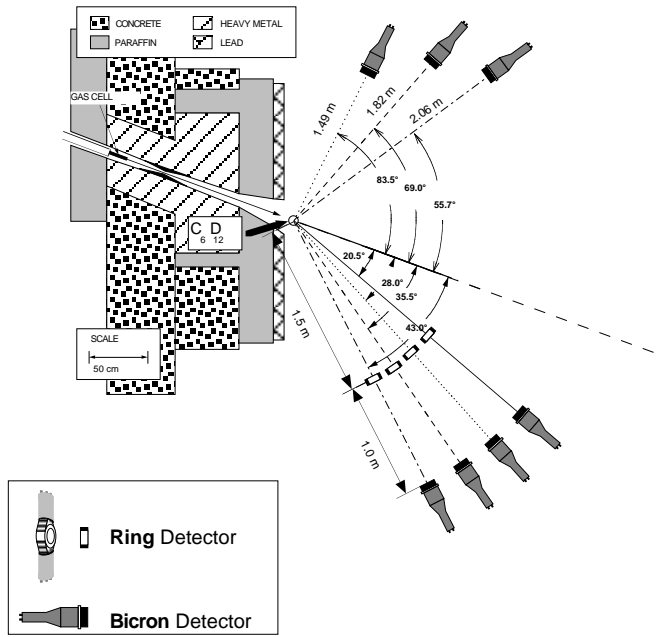,width=\linewidth}
  \end{center}
  \caption{The experimental setup for the {\em NN} FSI cross-section 
   measurements
   in {\em nd} breakup. All detectors are in the horizontal plane.  
   The two detectors used to measure the {\em np} FSI cross section at a 
   specific {\em np} angle pair lie on lines of the same type.
   The {\em nn} FSI cross sections were measured at each angle with the
   pair of ring and cylindrical detectors positioned at that angle.}
  \label{fig:1}
 \end{minipage}
\end{figure}

All measurements were
made  using the shielded neutron source at the Triangle Universities
Nuclear Laboratory (TUNL).  The experimental setup is shown in 
Fig.~\ref{fig:1}.  The
momentum of the two emitted neutrons and the energy of the proton 
in each breakup event were
measured, thereby overdetermining the kinematics.  The neutrons were
detected in liquid organic scintillators.  The energy of the emitted 
proton was measured in the deuterated
liquid scintillator
${\rm C}_6{\rm D}_{12}$ (NE-232), which served as the deuteron scatterer
and will be referred to as the central
detector (CD).  The energies of the
outgoing neutrons were determined by measuring their flight times
from the CD to the neutron detectors. The neutron detectors on the right
side of the incident beam axis were used for the {\em nn} FSI
measurements.  At each angle of the {\em nn} pair, one neutron was 
detected in the ring-shaped detector placed 1.5~m from the CD
and the other in the coaxial
cylindrical detector placed 2.5~m from the CD, which was
positioned to fill the solid 
angle of the opening in the ring detector.

In the {\em np} FSI one neutron moves in the same direction
as the proton, and the other one is emitted on the opposite
side of the incident neutron-beam axis. 
The neutrons emitted on the right side were detected in either the
ring-shaped detector or in the cylindrical detector. 
The associated neutrons were emitted to the left side and were detected
in cylindrical scintillators located as shown in Fig.~\ref{fig:1}.  
All neutron detectors were filled with a liquid scintillator fluid 
with $n-\gamma$
pulse-shape sensitivity (either NE213 or BC501A).  The active volume
of each cylindrical
detector was 12.6~cm dia. $\times$ 5.5~cm thick, and that of each ring
detector was 7.6~cm inner dia. $\times$ 13.4~cm outer dia. $\times$
4.0~cm thick.
The neutron detector efficiencies were determined in a dedicated
series of measurements using neutrons from the 
$^2{\rm H}(d,n)^3{\rm He}$ reaction
and from a $^{252}$Cf source.  The energy dependence of the relative
detection efficiency of each detector was
determined to an accuracy of $\pm{1.0}\%$, and the absolute efficiency 
to $\pm{2.5}\%$.

The neutron beam was produced using the $^2{\rm H}(d,n)^3{\rm He}$
reaction.
The production target was a 3-cm long cell pressurized with 7.8 atm
of deuterium gas.  The cell was bombarded with a
10.0-MeV d.c. deuteron beam, which
entered the cell through a 6.35-$\mu$m thick Havar containment
foil and was stopped in a gold end cap.  The neutron energy spread
was 400 keV. 
The detector area was
shielded from the neutron production target by a 1.7-m thick
multiple component wall.
The neutron beam at the CD was defined by a rectangular double-truncated
collimator, which was designed such that the CD was illuminated
almost exclusively by unscattered neutrons produced in the deuterium
gas cell.  The deuteron
beam current on target was about $2\:\mu{A}$, and the counting rate of
the CD was about 400~kHz with a threshold setting of one-tenth of the 
Compton-scattering edge for $\gamma$ rays from a $^{137}$Cs source.  
The rate in the CD electronics set the limit on the maximum acceptable
beam current.  The pulse-height thresholds on the neutron detectors were set 
at one-third of the $^{137}Cs$ Compton-scattering edge.  Data
were collected for a total of 2000 hours.

The integrated beam-target luminosity was determined by measuring the yields
for {\em nd} elastic scattering concurrently with the data from the breakup
reaction.  Since the differential cross section for {\em nd} elastic 
scattering can be calculated using realistic {\em NN} potentials in the 
Faddeev method with high numerical precision \cite{Glo96} and the
calculations agree well with existing data, we 
elected to use calculated cross sections rather than experimental data 
in the luminosity determination.
The simultaneous accumulation of the breakup and
{\em nd} elastic-scattering data reduced the
sensitivity of our measurements to system deadtimes and
absolute detection efficiencies.  Also this technique eliminated the need for 
absolute measurements of the neutron flux
and the thickness of the CD. 

The width of the coincidence window in the event-trigger circuit
was 400~ns and allowed for the 
concurrent measurements
of {\em true} breakup events and events due to the {\em accidental}
coincidences between signals from the CD and the neutron detectors. 
All events that satisfied conservation of energy within $\pm$2~MeV 
were projected into 0.5-MeV wide bins
along the point-geometry kinematic locus.  The distance
along the kinematic curve of $E_{n1}$ versus $E_{n2}$, where $E_{n1}$ and
$E_{n2}$ are the energies of the two emitted neutrons, will be referred to
as {\em S}.  The value of {\em S} is set to zero 
at the point where $E_{n2} = 0$ and increases as one moves 
counterclockwise around the locus.  The {\em true+accidental} and 
{\em  accidental} events were projected onto the ideal locus separately.
The projection was done using the
minimum-distance technique described by Finckh {\em et al.} \cite{Fin87}.
The {\em true} events were obtained by subtracting {\em accidental} from 
{\em true+accidental} events.

The measured cross sections were compared to Monte-Carlo (MC) simulations
that included the energy resolution and the finite geometry of the
experimental setup. 
The basis of these simulations was theoretical point-geometry cross-section 
libraries generated for a range of $a_{np}$ ($a_{nn}$) values at incident 
neutron energies of 12.8, 13.0 and 13.2 MeV. 
The point-geometry cross sections were 
obtained from transition matrix elements of the breakup operator $U_0$ 
\begin{eqnarray}
 U_{0} &=& (1 + P)\tilde T,
\label{eq3}
\end{eqnarray} 
where the $\tilde T$ operator 
sums up all multiple scattering contributions through the three-nucleon (3N) 
Faddeev integral 
equation
\begin{eqnarray}
\tilde T{\vert}{\phi}> &=& tP{\vert}{\phi}> + 
(1+tG_0)V_4^{(1)}(1+P){\vert}{\phi}> 
+ tPG_0\tilde T{\vert}{\phi}> \nonumber \\ 
 &+& (1+tG_0)V_4^{(1)}(1+P)G_0\tilde T{\vert}{\phi}>.
\label{eq4}
\end{eqnarray}
Here $G_0$ is the free 3N propagator, {\em t} is the {\em NN} t-matrix, and 
operator {\em P} is the sum 
of a cyclical and anticyclical permutation of three nucleons. In the generation
of the libraries, the terms containing $V_4$, the 3NF potential, were set
to zero.

The cross-section libraries were obtained using the Bonn-B (OBEPQ) {\em NN} 
potential \cite{Mac89}. This potential is fitted in the $^1S_0$ state to the 
experimentally-determined value of $a_{np}$. The charge-indepence breaking in 
the $^1S_0$ {\em NN} force is imposed by using for the $^1S_0$ {\em nn} force 
the version of the Bonn-B potential \cite{Mac89} that was fitted to {\em pp} 
data. To account for charge-symmetry breaking in the calculations, the total 
3N isospin T=3/2 admixture has been included~\cite{chdep}. 
For the purpose of this analysis, modifications of 
the  $^1S_0$ {\em NN} interaction were accomplished 
by adjusting the $\sigma$-meson coupling constant 
$g_{\sigma}^2/4\pi$ \cite{Mac89}. In this way 
$^1S_0$ {\em np} ({\em nn}) interactions with different {\em np} ({\em nn})
scattering lengths were generated.

\begin{figure}
 \noindent
 \begin{minipage}{\linewidth}
  \begin{center}
   \epsfig{file=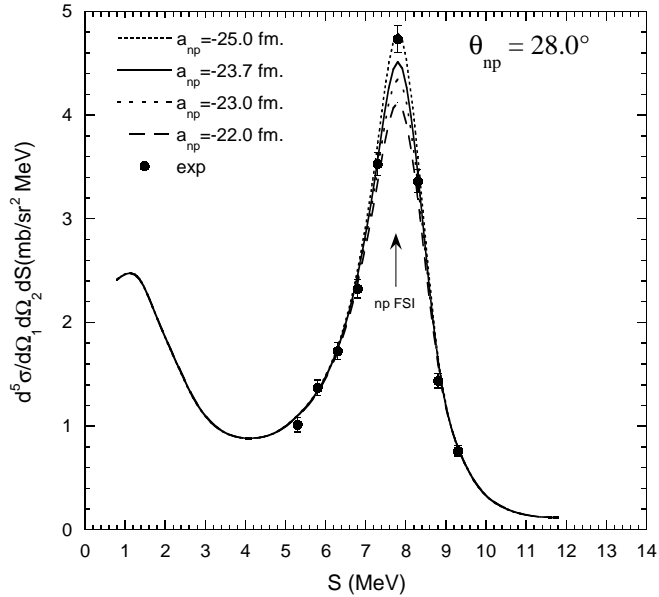,width=\linewidth,clip=}
  \end{center}
  \caption{Cross sections for $\theta_{np} = 28.0^\circ$ 
   ($\theta_{n1}$ = $28.0^\circ$, $\theta_{n2}$= $83.5^\circ$, 
     $\phi_{12}$ = $180^\circ$).  
   The points are the data from the present work.  The curves are MC 
   simulations based on {\em nd} calculations
   made with four values of $a_{np}$: -22.0, -23.0, -23.75 and -25.0 fm.}
  \label{fig:2}
 \end{minipage}
\end{figure} 

\begin{figure}
 \noindent
 \begin{minipage}{\linewidth}
  \begin{center}
   \epsfig{file=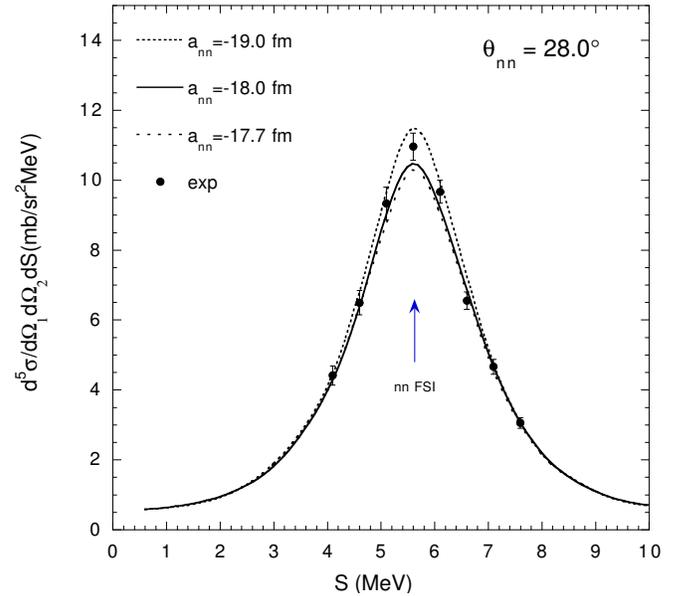,width=\linewidth,clip=}
  \end{center}
  \caption{Cross~sections~for~$\theta_{nn}=28.0^\circ$ 
   ($\theta_{n1}$ = $\theta_{n2}$ = $28.0^\circ$, $\phi_{12}$ = $0^\circ$).  
   The points are the data from the present work.  The curves are MC 
   simulations based on {\em nd} calculations
   made with three values of $a_{nn}$: -17.7, -18.0, and -19.0 fm.}
  \label{fig:3}
 \end{minipage}
\end{figure}

Simulated cross sections in comparison to our data for 28.0$^\circ$ 
are  shown in Figs.~\ref{fig:2} and \ref{fig:3} for several values
of $a_{np}$ and $a_{nn}$, respectively.  
A value of $a_{NN}$ and its statistical uncertainty were 
determined for each detector-pair configuration using a single-parameter 
(either $a_{np}$ or $a_{nn}$) minimum-$\chi^2$ fit to the 
absolute cross-section data.
The results are given in Table~\ref{tab:1}. The uncertainties listed in 
Table~\ref{tab:1}
are statistical only.
The systematic uncertainties in our determinations
are $\pm$0.8~fm for $a_{np}$ and $\pm$0.6~fm for $a_{nn}$.
Uncertainties in the
neutron detector efficiencies and the integrated target-beam
luminosity account for about 80\% of the systematic uncertainty.  

We observed no signficant angle dependence in $a_{np}$, and the
consistency in the $a_{nn}$ data from one angle to the next is 
statistically acceptable. 
Combining the statistical and systematic uncertainties in quadrature,
we obtain $a_{np}$ = -23.5 $\pm$ 0.8~fm and $a_{nn}$ = -18.7 $\pm$ 0.6~fm.  
Our result for $a_{np}$ is in agreement with the value of $a_{np}$
($-23.748{\pm}0.009$~fm \cite{Koe75}) obtained from free {\em np}
scattering measurements.
We use this result to set an upper limit on the 
influence of 3NF on the value of {\em NN} scattering
lengths determined from our experiment, i.e.,
$\Delta a^{3NF}_{np} \le a_{np}^{nd} - a_{np}^{free} = 0.2 \pm 0.8$~fm,
where {\em free} and {\em nd} refer to values obtained 
from data for free {\em np} scattering and
from {\em nd} breakup, respectively.  
This result is consistent with zero. 
Scaling our
result for $a_{np}$ by the ratio of $a_{nn}$ to $a_{np}$, we obtain 
the upper limit 
due to 3NF effects in the {\em nd} breakup reaction to be  
$\Delta a^{3NF}_{nn} 
\le 0.2 \pm 0.6$~fm, which is also consistent with zero.  

Using the TM 3NF model we estimated possible effects of 3NF on {\em np} FSI 
cross sections. 
The calculations were produced by
solving Eq.(\ref{eq4}) using four 
modern {\em NN} potentials: AV18~\cite{Wir95}, CD Bonn~\cite{cdbonn}, 
NijmI and NijmII~\cite{nijm}. In each calculation the 
2$\pi$-exchange TM 3NF potential~\cite{tm3nf}
was included as $V_4$, which was split into 3 parts  
where each one was symmetrical under exchange of two particles. 
For instance, for  
the 2$\pi$-exchange 3NF~\cite{tm3nf}, this corresponds 
to the three possible choices of the nucleon which undergoes the (off-shell) 
$\pi-N$ scattering. In our calculations with the 3NF the 
strong cut-off parameter  $\Lambda$ in the TM 3NF model was adjusted 
separately for each 
{\em NN} potential to reproduce the experimental triton 
binding energy~\cite{nog97}.  For details 
of the formalism and the numerical treatment refer to
refs.~\cite{Glo96,Actap}. 
We calculated the percentage deviation caused by effects
of the TM 3NF 
on the {\em np} FSI 
point-geometry  cross section as the function of $\theta_{np}$ 
at an incident neutron energy of 13 MeV. At all angles and for
all potentials the change in the cross section due to the addition
of the TM 3NF never exceeds 6\%. 
For the {\em np} FSI production angles of the present experiment it is  
limited to the range of 1\% to 4\%, which corresponds to a theoretical
range of $(\Delta a^{3NF}_{np})_{th}$ from -0.8 to -0.2 fm. 
Our experimentally determined $\Delta a^{3NF}_{np}$ is within two
standard deviations of the predictions using any of the four {\em NN}
potentials with the TM 3NF adjusted to fit the triton binding energy.

\begin{center}
\begin{minipage}{\linewidth}
\begin{table}
 \noindent
  \caption{The $a_{np}$ and $a_{nn}$ values extracted from the 
    fit to the present data for the absolute {\em NN} FSI cross section in 
    {\em nd} breakup at the angles measured in the present study. The 
    $\chi^2$ per datum for the best is given at each angle.  The weighted 
    mean of the data for all angles is given in the bottom row.  All 
    uncertainties are statistical only.}  
\begin{tabular}{|c|c c||c c|} 
    $\theta_{NN}$ & $a_{np}$ (fm) & $\chi^2$/pt & 
 $a_{nn}$ (fm) & $\chi^2$/pt \\ \hline
      $43.0^\circ$ & -23.6 $\pm$ 0.3 & 2.1 & -18.8 $\pm$ 0.4 & 1.5 \\ 
      $35.5^\circ$ & -23.2 $\pm$ 0.3 & 2.7 & -17.7 $\pm$ 0.4 & 0.6 \\
      $28.0^\circ$ & -23.7 $\pm$ 0.3 & 3.0 & -18.8 $\pm$ 0.2 & 0.1 \\ 
      $20.5^\circ$ &                 &     & -18.9 $\pm$ 0.2 & 0.8 \\ \hline
{\bf Mean} & -23.5 $\pm$ 0.2 &	2.6 & -18.7 $\pm$ 0.1 & 0.6 \\ 
\end{tabular}
\label{tab:1}
\end{table}
\end{minipage}
\end{center}  

Summarizing, our measured values are $a_{np}$ = -23.5 $\pm$0.8~fm and
$a_{nn}$ = -18.7 $\pm$0.6~fm. 
In our analysis magnetic interactions are not considered.  Their
possible effects~\cite{Slo84} have to be studied.
By comparing our results for $a_{np}$ to
the recommended value from {\em np} free scattering and scaling by the ratio
of $a_{nn}$ to $a_{np}$, we set an upper limit of 
$\Delta a^{3NF}_{nn} = 0.2 \pm 0.6$~fm on the contribution of 
3NF effects on our value of $a_{nn}$. Though the experimental results
suggest an opposite sign for $\Delta a^{3NF}_{np}$ than predicted using 
the TM 3NF, our value is
consistent with the theoretical predictions
within the reported uncertainties.
Since our value for $a_{np}$ obtained from {\em nd}
breakup 
agrees with that from free {\em np} scattering, we conclude that our
investigation of the {\em nn} FSI done under identical conditions
should lead to a valid measure of $a_{nn}$.  Our value for $a_{nn}$ is in 
agreement with the recommended value~\cite{Mil90}, which
comes from $\pi^-d$ capture measurements, and disagrees with values 
obtained from earlier {\em nd} breakup studies~\cite{Sla89}.

\acknowledgments
This work was supported in part by the U.S. Department of Energy,
Office of High Energy and Nuclear Physics, under grant No.
DE-FG02-97ER41033, by the Maria Sklodowska-Curie II Fund under grant
No. MEN/NSF-94-161, by the USA-Croatia NSF Grant JF129, and by the European
Community Contract No. CI1*-CT-91-0894.  The numerical calculations
were performed on the Cray T916 of the North Carolina Supercomputing
Center at the Research Triangle Park, North Carolina and on the Cray
T90 and T3E of the H\"oechstleistungsrechenzentrum in J\"ulich, Germany.

\end{multicols}
\end{document}